\documentclass[floatfix,showpacs,aps,nofootinbib]{revtex4}
\usepackage{amssymb}

\usepackage[dvips]{epsfig}
\usepackage[english]{babel}

\usepackage[utf8]{inputenc}
\usepackage{array}
\usepackage{amsmath}
\usepackage{amsfonts}
\usepackage{amssymb}
\usepackage{xcolor}
\usepackage{hyperref}
\hypersetup{colorlinks=true,linkbordercolor=blue,linkcolor=blue, citecolor=blue}
\usepackage{subeqnarray}
\usepackage{tensor}
\usepackage{indentfirst} 
\newcommand{\lambdabp}{\stackrel{\neg}{\lambda}_{h}(\boldsymbol{p})}
\newcommand{\lambdap}{\lambda_{h}(\boldsymbol{p})}

\newcommand{\gm}{\gamma_{\mu}}
\makeatletter
\renewcommand*\l@section{\@dottedtocline{1}{1.5em}{2.3em}}
\makeatother

\begin{document}

\title{On the bilinear covariants associated to mass dimension one spinors}

\author{ J. M. Hoff da Silva\footnote{hoff@feg.unesp.br}}
\affiliation{UNESP - Campus de Guaratinguetá - DFQ - Avenida Dr. Ariberto Pereira da Cunha, 333 - CEP 12516-410 - Guaratinguetá - SP - Brazil.}

\author{C. H. Coronado Villalobos\footnote{ccoronado@feg.unesp.br}}
\affiliation{UNESP - Campus de Guaratinguetá - DFQ - Avenida Dr. Ariberto Pereira da Cunha, 333 - CEP 12516-410 - Guaratinguetá - SP - Brazil.}

\author{R. J. Bueno Rogerio\footnote{rodolforogerio@feg.unesp.br}}
\affiliation{UNESP - Campus de Guaratinguetá - DFQ - Avenida Dr. Ariberto Pereira da Cunha, 333 - CEP 12516-410 - Guaratinguetá - SP - Brazil.}

\author{E. Scatena\footnote{e.scatena@ufsc.br}}
\affiliation{Universidade Federal de Santa Catarina - CEE - Rua Pomerode, 710 - CEP 89065-300 - Blumenau - SC - Brazil.}

\date{\today}

\begin{abstract}
In this paper we approach the issue of Clifford algebra basis deformation, allowing for bilinear covariants associated to Elko spinors which satisfy the Fierz-Pauli-Kofink identities. We present a complete analysis of covariance, taking into account the involved dual structure associated to Elko. Moreover, the possible generalizations to the recently presented new dual structure are performed. 
\end{abstract}

\maketitle

\section{Introduction}
The so-called Elko spinors are a new set of spinors with a complex and interesting structure on its own. Historically, they were proposed by Ahluwalia and Grumiller when studying properties of the Majorana spinor. Similar to them, Elko spinors are eingenspinors of the charge conjugation operator, $C$, but they have dual helicity and can take positive (self-conjugated) and negative (anti-self-conjugated) eigenvalues of $C$, whilst the Majorana ones take only the positive value and carry single-helicity.

From the physical point of view, Elko spinors are constructed to be invisible to the other particles (i.e., it does not couple with the fields of the Standard Model, except for the Higgs boson), becoming a natural candidate to dark matter \cite{jcap}. Mathematically, the dual helicity peculiarity forces to a redefinition of the dual spinor structure, as can be seem in \cite{jcap}. This idiosyncrasy reflects itself when constructing the spin sums for the Elko spinors, which breaks Lorentz symmetry. However, the spin sums are invariant under transformations of the Very Special Relativity (VSR) \cite{horvath, cohen, ahluwaliahorvath}. There are several areas in which Elko spinors have been studied, from accelerator physics \cite{fen1,fen2,fen3,fen4} to cosmology \cite{co1,co2,co3,co4,co5,co6,co7,co8,co9,co10,co11}. In particular, the appreciation of new dual structures bring interesting possibilities within the algebraic scope \cite{cjr,rcd}. 

As it is well known, much of the physics associated to spinor fields is unveiled from its bilinear covariants by the simple reason that single fermions are not directly experienced. In this context it is indeed important to pay special attention to the subtleties of Clifford algebra when associating real numbers to the bilinear covariants \cite{lounesto}. It may sound as an secondary issue, but in fact the opposite is true. In two outstanding papers in the ninetieths \cite{crawford1,crawford2}, Crawford worked out several important formalizations concerning the bispinor algebra. Among these, a rigorous procedure to obtain real bilinear covariants was developed. The general aim of this paper is to make use of this procedure to envisage what (if any) bilinear covariants are real when dealing with mass dimension one spinors. To the best of our hope, the results to be shown here may shed some light on the observables associated to these spinors. With suitable, but important, changes we take advantage of the formalism developed in \cite{crawford1} in order to study the bilinear covariants associated to the Elko spinor case. After a complete analysis, including the right observance of the Fierz-Pauli-Kofink (FPK) \cite{FPK, beres} relations, we arrive at the subset of real bilinear covariants. 

Quite recently, new possibilities concerning the field adjoint possibilities was investigated in deep \cite{1305,nosso}. These formalizations may lead to a local and full Lorentz spin $1/2$ field also endowed with mass dimension one, evading, thus, the Weinberg's no-go theorem \cite{Ahluwa2}. We also have investigated the bilinears to this case, and in some extent the aforementioned program may be applied, leading to similar conclusions.  

This paper is organized as follows: in Section II we introduce the bilinears covariant and proceed with the Clifford algebra basis deformation then, in Section III we analyze its covariant structure. Both these sections are related to Elko spinors as objects whose spin sums break Lorentz symmetry. The natural background symmetry is, then, encoded into the orthochronous proper Lorentz subgroup, leaving for the Appendix B an analogous investigation taking into account VSR Elko spinors. In Section IV, we conclude with some remarks about the results we found.
    
\section{Bilinear analysis}

Let $\psi$ be a given spinor field belonging to a section of the vector bundle $\mathbf{P}_{Spin^{e}_{1,3}}(\mathcal{M})\times\, _{\rho}\mathbb{C}^4$ where $\rho$ stands for the entire representation space $D^{(1/2,0)}\oplus D^{(0,1/2)}$, or a given sector of such. The bilinear covariants associated to $\psi$, as usual, reads
\begin{eqnarray}
\label{covariantes}
\sigma=\psi^{\dag}\gamma_{0}\psi, \hspace{1cm}  \omega=-\psi^{\dag}\gamma_{0}\gamma_{0123}\psi, \hspace{1cm} \mathbf{J}=\psi^{\dag}\mathrm{\gamma_{0}}\gamma_{\mu}\psi\;\gamma^{\mu}, \nonumber\\
\mathbf{K}=\psi^{\dag}\mathrm{\gamma_{0}}\textit{i}\mathrm{\gamma_{0123}}\gamma_{\mu}\psi\;\gamma^{\mu},\hspace{1cm} \mathbf{S}=\frac{1}{2}\psi^{\dag}\mathrm{\gamma_{0}}\textit{i}\gamma_{\mu\nu}\psi\gamma^{\mu}\wedge\gamma^{\nu}\,,
\end{eqnarray} where the Dirac matrices are written in the Chiral (or Weyl) representation
\begin{eqnarray}
\gamma_0 = \left( \begin{array}{cc}
\mathbb{O} & \mathbb{I} \\ 
\mathbb{I} & \mathbb{O}
\end{array}  \right), \qquad \gamma_i = \left( \begin{array}{cc}
\mathbb{O} & \sigma_{i} \\ 
-\sigma_{i} & \mathbb{O}
\end{array}  \right).
\end{eqnarray}

In general grounds, it is always expected to associate (\ref{covariantes}) to physical observables. For instance, in the usual case, bearing in mind the relativistic description of the electron, $\sigma$ the invariant length, $\boldsymbol{J}$ is associated to the current density, $\boldsymbol{K}$ is the spin projection in the momentum direction, and $\boldsymbol{S}$ is the momentum electromagnetic density. The bilinear covariants, as well known, obey the so-called Fierz-Pauli-Kofink (FPK) identities, given by \cite{baylis}
\begin{eqnarray}\label{fpkidentidades}
\boldsymbol{J}^2 = \sigma^2+\omega^2, \quad J_{\mu}\!\!&K_{\nu}&\!\!-K_{\mu}J_{\nu} = -\omega S_{\mu\nu} - \frac{\sigma}{2}\epsilon_{\mu\nu\alpha\beta}S^{\alpha\beta}, \nonumber
\\
J_{\mu}K^{\mu} &=& 0, \quad \boldsymbol{J}^2 = -\boldsymbol{K}^2.
\end{eqnarray}
It can be seen that the physical requirement of reality can always be satisfied for Dirac spinors bilinear covariants \cite{crawford1}, by a suitable deformation of the Clifford basis leading to physical appealing quantities. Unfortunately, the same cannot be stated for mass dimension one spinors, as Elko spinors. Actually, a straightforward calculation shows an incompatibility in the usual construction of bilinear covariants. In fact, one of the FPK identities is violated. The reason rests upon the new dual structure associated to these spinors. It is worth to mention that the main difference between the Crawford deformation \cite{crawford1, crawford2} and the one to be accomplished here is that in the former case, the spinors are understood as Dirac spinors, i. e., spinorial objects endowed with single helicity. Therefore, the dual structure is the usual one $\bar{\psi}(\boldsymbol{p}) = \psi^{\dag}(\boldsymbol{p})\gamma_0$ and the required normalization is also ordinary. On the other hand Elko spinors, due to its own formal structure, need a dual redefinition $\stackrel{\neg}{\lambda}^{S/A}_{h}(\boldsymbol{p}) = [\Xi(\boldsymbol{p})\lambda^{S/A}_{h}(\boldsymbol{p})]^{\dag}\gamma_0$. This redefinition leads, ultimately, to a new normalization culminating in a basis deformation satisfying the FPK identities.

Let us make these assertions more clear by explicitly showing the mentioned problem. In order to guarantee the sequential readability of the paper we leave for the Appendix A a brief, but self contained, overview on the spinorial formal structure of Lorentz breaking Elko fields. Taking advantage of what was there defined, we use as an example the spinor $\lambda^{S}_{\lbrace -,+\rbrace}(\boldsymbol{p})$ and its dual given by, respectively
\begin{eqnarray}\label{3}
\lambda^{S}_{\lbrace -,+\rbrace}(\boldsymbol{p}) = \Upsilon_{-} 
\left(\begin{array}{c}
-i\sin\theta/2 e^{-i\phi/2} \\ 
i\cos\theta/2 e^{i\phi/2} \\  
\cos\theta/2 e^{-i\phi/2} \\ 
\sin\theta/2 e^{i\phi/2}
\end{array}  \right),
\end{eqnarray}   
and
\begin{eqnarray}\label{14}
\stackrel{\neg}{\lambda}^{S}_{\lbrace -,+\rbrace}(\boldsymbol{p}) =\Upsilon_{+}\left( \begin{array}{cccc}
-i\sin\theta/2 e^{i\phi/2} & i\cos\theta/2 e^{-i\phi/2} & -\cos\theta/2 e^{i\phi/2} & -\sin\theta/2 e^{-i\phi/2}
\end{array}  \right)\!.
\end{eqnarray} We reinforce once again that the dual structure associated to Elko spinors are obtained in a very judicious fashion \cite{1305}, leaving no space to modifications, exception made to the generalizations found in Ref. \cite{jcap,1305}. Using (\ref{3}) and (\ref{14}), as a direct calculation shows, Eqs. (\ref{covariantes}) give 
\begin{eqnarray}
\sigma &=& -2m,\label{cov1}\\
\omega &=& 0,\label{cov2}\\
J_0 &=& 0, \nonumber\\
J_1 &=& 2im\cos\theta\cos\phi,\nonumber \\
J_2 &=& 2im\cos\theta\sin\phi,\nonumber\\
J_3 &=& -2im\sin\theta, \label{cov3}\\
K_0 &=& 0,\nonumber\\
K_1 &=& -2m\sin\phi,\nonumber\\
K_2 &=& 2m\cos\phi,\nonumber\\
K_3 &=& 0,\label{cov4}
\end{eqnarray}
and 
\begin{eqnarray}
S_{01} &=& -2im\sin\theta\cos\phi,\nonumber\\
S_{02} &=& -2im\sin\theta\sin\phi,\nonumber\\
S_{03} &=& -2im\cos\theta,\nonumber\\
S_{12} &=& S_{13} = S_{23} = 0.\label{cov5}
\end{eqnarray}
As it can be verified, the above bilinear covariants do not obey the FPK equations. More specifically, the relation containing $S_{\mu\nu}$ is not satisfied. In view of this problem, we revisited the formulation performed in \cite{crawford1} in order to find out an appropriated Clifford basis upon which the bilinear covariants can be constructed, leading to the right verification of the FPK relations. The price to be paid is that only a subset of bilinear covariants comprises real quantities. 

\subsection{Deformation of the Clifford algebra basis}

As it is well known, the very constitutive relation of the Clifford algebra is given by
\begin{eqnarray}\label{eita}
\lbrace \gamma_{\mu}, \gamma_{\nu}\rbrace = 2g_{\mu\nu}\mathrm{I}, \quad \mu,\nu = 0,1,2,...,N-1,
\end{eqnarray}
where $g_{\mu\nu}$ is a $N=2n$ even-dimensional space-time metric, which in Cartesian coordinates has the form $diag(1,-1,...,-1)$. The generators of the Clifford algebra are, then, the identity $\mathrm{I}$ and the vectors $\gamma_\mu$, usually represented as square matrices. The standard approach dictates the complementation of the Clifford algebra basis, in order to guarantee real bilinear covariants. This complement is performed by the composition of the vector basis, used as building blocks \cite{crawford1} 
\begin{eqnarray}\label{gammatil}
\tilde{\gamma}_{\mu_{1}\mu_{2}...\mu_{N-M}} \equiv \frac{1}{M!}\epsilon_{\mu_{1}\mu_{2}...\mu_{N}}\gamma^{\mu_{N-M+1}}\gamma^{\mu_{N-M+2...}}\gamma^{\mu_{N}}.
\end{eqnarray} 
As it is easy to see, the lowest $M$ value is two (the smallest combination), nevertheless, it runs in the range $M=2,3,...,N$. In this respect, the elements that form the (real) Clifford algebra basis are 
\begin{eqnarray}\label{set}
\lbrace \Gamma_{i}\rbrace \equiv \lbrace \mathrm{I}, \gm, \tilde{\gamma}_{\mu_{1}\mu_{2}...\mu_{N-2}},..., \tilde{\gamma}_{\mu},\tilde{\gamma}\rbrace,  
\end{eqnarray} where $\tilde{\gamma}\equiv \tilde{\gamma}_{\mu_{1}\mu_{2}...\mu_{N-N}}$.    

In view of the new elements appearing in the definition of the Elko dual, it is necessary to adequate the Clifford algebra basis complementation. As shown previously, it is absolutely necessary for the right appreciation of the FPK relations. We shall stress that for the Dirac spinorial case, the set (\ref{set}) is suitable deformed (by a slightly different normalization) in order to provide real bilinear covariants. We shall pursue something similar here, and we are successful in correction the problem related to the FPK relations. Notwithstanding, only a subset of bilinear covariants ends up real in the Elko spinorial case.  

The first two bilinear arising from the Clifford algebra basis are 
\begin{eqnarray}
\sigma &\equiv&\lambdabp\mathrm{I}\lambdap,\label{sig}\\
J_{\mu} &\equiv&\lambdabp\gm\lambdap\label{jota},
\end{eqnarray}
where $\lambdabp =  [\Xi(\boldsymbol{p})\lambdap]^{\dag}\gamma_0$ is identified as the Elko dual spinor\footnote{We once again refers the reader to the Appendix A for the definition of the dual, as well as its necessity.}. The operator $\Xi(\boldsymbol{p})$ is responsible to change the spinor ``helicity'' (better said, the type of Elko spinor), here labeled by $h$. All the details concerning such an operator can be found in Ref. \cite{1305}. The requirement $\sigma=\sigma^\dagger$ leads automatically to $\gamma_0 = \Xi^{\dag}(\boldsymbol{p})\gamma_0^{\dag}\Xi(\boldsymbol{p})$ since $\Xi^{2}(\boldsymbol{p})=1$. This constraint is readily satisfied, in such a way that (\ref{sig}) is real\footnote{Notice that it must be proved that the bilinear composed with $\Xi(\boldsymbol{p})$'s are covariants. We shall investigate this issue in the next section.}. By the same token, one should require $J_{\mu} = J_{\mu}^{\dag}$. This imposition leads to the constraint $\gamma_0\gm = \Xi^{\dag}(\boldsymbol{p})\gm^{\dag}\gamma_0^{\dag}\Xi(\boldsymbol{p})$ which, however, cannot be fulfilled. This is an important point. In fact, the counterpart associated to Dirac spinors is simply $\gamma_0^{-1}\gamma_{\mu}^\dagger\gamma_0=\gamma_\mu$, a constraint naturally achieved. The new dual structure, therefore, has forced a new interpretation of the bilinear covariants. Firstly, as a matter of fact, for mass dimension one spinors $J^\mu$ cannot be associated to the conserved current. Obviously, in order to have $\partial_\mu J^\mu=0$ it is necessary the use of the Dirac equation. This truism has lead to interesting algebraic possibilities \cite{cjr}, but the point to be emphasized here is that there is no problem in having a complex quantity related to the bilinear $J^\mu$. The additional important consequence of an imaginary $J^\mu$ is that in order to satisfy the FPK equations one must have $K^\mu$ or $S^{\mu\nu}$ also imaginary.  

Notice that, instead of the usual Crawford deformation, here we do not arrive at an entire real bilinear set. In fact, in trying to implement the full reality condition it is mandatory to change the building block of the Clifford basis $\gamma_\mu$. It would inevitably lead, however, to a change in the constitutive algebraic relation of the Clifford algebra (\ref{eita}). Therefore, this change must be excluded. It is important to emphasize, moreover, that even being willing to accept a modification of (\ref{eita}) the resulting constraint to get a real set of bilinears cannot be fulfilled.

Having said that, we may proceed deforming the usual basis in order to redefine bilinear covariants which satisfy the FPK identities. Making use of Eq. \eqref{gammatil} and considering that the norm for the Elko spinor is real we have 
\begin{eqnarray}
[\lambdabp\tilde{\gamma}_{\mu_{1}\mu_{2}...\mu_{N-M}}\lambdap]^{\dag}= (-1)^{M(M-1)/2}\lambdabp\Xi(\boldsymbol{p})\tilde{\gamma}_{\mu_{1}\mu_{2}...\mu_{N-M}}\Xi(\boldsymbol{p})\lambdap. \label{urru}
\end{eqnarray} It can be readily verified that the following redefinition of $\tilde{\gamma}_{\mu_{1}\mu_{2}...\mu_{N-M}}$ is appropriate to ensure $K^\mu$ as a real quantity:
\begin{eqnarray}\label{gammatil2}
\tilde{\gamma}_{\mu_{1}\mu_{2}...\mu_{N-M}}= (i^{M(M-1)/2}/M!)\Xi(\boldsymbol{p})\epsilon_{\mu_{1}...\mu_{N}}\gamma^{\mu_{N-M+1}...}\gamma^{\mu_{N}}\Xi(\boldsymbol{p}).
\end{eqnarray}
With the redefinition above, one is able to define the bispinor Clifford algebra basis as in (\ref{set}), but with the gammas given by (\ref{gammatil2}). As an example, consider the four-dimensional space-time. In this case the basis is given, accordingly, by 
\begin{eqnarray}
M &=& 4 \quad\Rightarrow\quad \tilde{\gamma} = -i \Xi(\boldsymbol{p})\gamma_5\Xi(\boldsymbol{p}), \\
M &=& 3 \quad\Rightarrow\quad \tilde{\gamma}_{\mu} = -\Xi(\boldsymbol{p})\gamma_5\gamma_{\mu}\Xi(\boldsymbol{p}),\\
M &=& 2 \quad\Rightarrow\quad  \tilde{\gamma}_{\mu\nu} = \frac{i}{2}\Xi(\boldsymbol{p}) \gamma_{\mu}\gamma_{\nu} \Xi(\boldsymbol{p}),
\end{eqnarray}
where $\gamma_5=-i\gamma_0\gamma_1\gamma_2\gamma_3$. Now, with the real Clifford algebra basis at hands, it is possible to construct the bilinear forms, given by
\begin{eqnarray}\label{Elkonewbilinear}
\mathrm{I}&\quad\Rightarrow\quad&\sigma_{E} = \lambdabp\mathrm{I}\lambdap, \nonumber \\
\gamma_{\mu}&\quad\Rightarrow\quad& J_{\mu_{E}}= \lambdabp\gamma_{\mu}\lambdap,\nonumber\\
\tilde{\gamma}&\quad\Rightarrow\quad& \omega_{E} = -i\lambdabp\gamma_5\lambdap,\\
\tilde{\gamma}_{\mu}&\quad\Rightarrow\quad& K_{\mu_{E}}= -\lambdabp\Xi(\boldsymbol{p})\gamma_5\gamma_{\mu}\Xi(\boldsymbol{p})\lambdap,\nonumber\\
\tilde{\gamma}_{\mu\nu}&\quad\Rightarrow\quad& S_{\mu\nu_{E}} = i\lambdabp \Xi(\boldsymbol{p})\gamma_{\mu}\gamma_{\nu}\Xi(\boldsymbol{p})\lambdap.\nonumber
\end{eqnarray}
From above construction, after a simple but tedious checking process, one ensures that the slight modifications of the bilinear covariants are enough to guarantee that the FPK identities \eqref{fpkidentidades} are satisfied. After all, we arrive at $\sigma$ and $K^\mu$ as real non null quantities. As remarked in the Introduction, to the best of our hope these quantities shall be considered in the determination of possible experimental outcomes of the Elko construction.

\section{Covariant structure}

So far we have worked out quantities defined as (\ref{Elkonewbilinear}) claiming that they must be faced as bilinear covariants. While they are bilinear quantities, their covariant structure must be evinced. All the issue is related to the (necessary) presence of the $\Xi(\boldsymbol{p})$ operator. 

Supposing that the Elko spinor belongs to a linear representation of the symmetry group in question, in such a way that seen in another frame the field undergoes a transformation as
\begin{eqnarray}\label{tl}
\lambda^{\prime S/A}_{h}(\boldsymbol{p^{\prime}}) = S(\Lambda) \lambda_{h}^{S/A}(\boldsymbol{p}).
\end{eqnarray} There is a Dirac-like operator that annihilates Elko \cite{1305} (no related to the field dynamics) given by
\begin{eqnarray}\label{eqdiracElko}
\big(\gamma_{\mu}p^{\mu}\Xi(\boldsymbol{p}) \pm m\big)\lambda^{S/A}_{h}(\boldsymbol{p})=0, 
\end{eqnarray} from which we shall investigate the covariance. Applying the transformation \eqref{tl} for the fields in the equation \eqref{eqdiracElko}, we find 
\begin{eqnarray}
\big(\gamma_{\mu}p^{\prime\mu}\Xi(\boldsymbol{p}) \pm m\big)\lambda^{\prime S/A}_{h}(\boldsymbol{p^{\prime}})=0.
\end{eqnarray}  
The momentum can be written as $p^{\mu}\leftrightarrow i\partial^{\mu}$ and the partial derivative transforms usually as $\partial^{\prime\mu} = \Lambda^{\mu}_{\;\;\;\beta}\partial^{\beta}$. Therefore, in order to ensure covariance of Eq. (\ref{eqdiracElko}) it is necessary the following behavior of the Dirac matrices and the $\Xi(\boldsymbol{p})$ operator, respectively
\begin{eqnarray}
 \gamma^{\prime}_{\beta} &=& S(\Lambda)\gamma_{\mu}S^{-1}(\Lambda)\Lambda^{\mu}_{\;\;\;\beta},\label{gamma1}\\
 \Xi^{\prime}(\boldsymbol{p}) &=& S(\Lambda)\Xi(\boldsymbol{p}) S^{-1}(\Lambda).\label{xi1}
\end{eqnarray} Equation (\ref{gamma1}) is the usual requirement to be imputed to the gamma matrices in order to achieve a covariant Dirac equation. The requirement (\ref{xi1}) is the new ingredient of the Elko theory, which must be investigated. 

Interestingly enough, from the expression \eqref{XI}, along with \eqref{gamma1}, it is possible to see that \cite{1305,speranca}
\begin{eqnarray*}
\Xi^{\prime}(\boldsymbol{p}) &=&\frac{1}{2m}\Big( \lambda^{\prime S}_{\lbrace +-\rbrace}(\boldsymbol{p^{\prime}})\bar{\lambda}^{\prime S}_{\lbrace +-\rbrace}(\boldsymbol{p^{\prime}}) + \lambda^{\prime S}_{\lbrace -+\rbrace}(\boldsymbol{p^{\prime}})\bar{\lambda}^{\prime S}_{\lbrace -+\rbrace}(\boldsymbol{p^{\prime}}) - \lambda^{\prime A}_{\lbrace +-\rbrace}(\boldsymbol{p^{\prime}})\bar{\lambda}^{\prime A}_{\lbrace +-\rbrace}(\boldsymbol{p^{\prime}})-\lambda^{\prime A}_{\lbrace -+\rbrace}(\boldsymbol{p^{\prime}})\bar{\lambda}^{\prime A}_{\lbrace -+\rbrace}(\boldsymbol{p^{\prime}}) \Big), \\
&=& \frac{1}{2m}S(\Lambda)\Big( \lambda^{S}_{\lbrace +-\rbrace}(\boldsymbol{p})\bar{\lambda}^{S}_{\lbrace +-\rbrace}(\boldsymbol{p}) + \lambda^{S}_{\lbrace -+\rbrace}(\boldsymbol{p})\bar{\lambda}^{S}_{\lbrace -+\rbrace}(\boldsymbol{p}) - \lambda^{A}_{\lbrace +-\rbrace}(\boldsymbol{p})\bar{\lambda}^{A}_{\lbrace +-\rbrace}(\boldsymbol{p})-\lambda^{A}_{\lbrace -+\rbrace}(\boldsymbol{p})\bar{\lambda}^{A}_{\lbrace -+\rbrace}(\boldsymbol{p}) \Big)\gamma_0 S^{\dag}(\Lambda)\gamma_0,
\end{eqnarray*} and therefore 
\begin{eqnarray}
\Xi^{\prime}(\boldsymbol{p}) = S(\Lambda)\Xi(\boldsymbol{p}) S^{-1}(\Lambda),
\end{eqnarray} 
as expected. Once verified the right transformations, we are able to evince the bilinear quantities. Starting from $\sigma$, we have
\begin{eqnarray*}
\sigma_{E}^{\prime} &=& \stackrel{\neg}{\lambda}^{\prime S/A}_{h}(\boldsymbol{p^{\prime}})\lambda^{\prime S/A}_{h}(\boldsymbol{p^{\prime}})\\
&=&\lambda^{\dag S/A}_h(\boldsymbol{p}) S^{\dag}(\Lambda)S^{-1\dag}(\Lambda)\Xi^{\dag}(\boldsymbol{p}) S^{\dag}(\Lambda)\gamma_0 S(\Lambda)\lambda^{S/A}_h(\boldsymbol{p}) \\
&=&  \stackrel{\neg}{\lambda}^{S/A}_{h}(\boldsymbol{p})\lambda^{S/A}_{h}(\boldsymbol{p}),\\
&=&  \sigma_{E},
\end{eqnarray*} 
implying $\sigma$ a scalar. Repeating the same procedure for the remaining bilinear forms, we obtain 
\begin{eqnarray*}
J^{\prime}_{\mu_{E}}&\rightarrow & \Lambda^{\nu}_{\mu}\lambdabp\gamma_{\nu}\lambdap, \quad \mbox{(Vector)},\\
\omega^{\prime}_{E} &\rightarrow &  det(\Lambda)i\lambdabp\Xi(\boldsymbol{p})\gamma_5\Xi(\boldsymbol{p})\lambdap, \quad  \mbox{(Scalar)},\\
K^{\prime}_{\mu_{E}}&\rightarrow & -det(\Lambda)\Lambda^{\nu}_{\;\;\;\mu}i\lambdabp\Xi(\boldsymbol{p})\gamma_5\gamma_{\nu}\Xi(\boldsymbol{p})\lambdap, \quad \mbox{(Vector)},\\
S^{\prime}_{\mu\nu_{E}} &\rightarrow & \frac{i}{2} \Lambda^{\alpha}_{\;\;\;\rho} \Lambda^{\beta}_{\;\;\;\vartheta}\lambdabp \Xi(\boldsymbol{p})\gamma_{\alpha}\gamma_{\beta}\Xi(\boldsymbol{p})\lambdap, \quad \mbox{(Bivector).}
\end{eqnarray*} 
Therefore, the nomenclature previously adopted is indeed adequate to the case at hands. We shall finalize pointing out that the investigation of covariance in which concerns $SIM(2)$, $HOM(2)$ Lorentz subgroups is taken into account in Appendix B. The analysis is quite analogous, and the physical statements are essentially the same. 

\section{Final Remarks}

In this paper we have shown that it is necessary to deform the usual Clifford algebra in order to ascertain the right observance of the Fierz-Pauli-Kofink identities, regarding Elko spinor fields. As a result, only a subset of bilinear covariants are real. After have found the proper deformation, and observed its dependence on the new dual operator, we study the covariance of the relevant quantities showing explicitly its behavior under a typical transformation of the relativistic group in question.  

We would like to finalize this paper by returning to the point raised at the end of the introductory Section. As already mentioned, in Refs. \cite{1305,Ahluwa2} it is described a subtle way to evade Weinberg's no-go theorem, by exploring another possibility of the dual structure, this time constructing the dual with the additional requirement of Lorentz invariant spin sums. Additional mathematical support was given in \cite{nosso}. In the formulation presented in \cite{1305}, it was taken a great care in the new dual structure in order not to jeopardize the simplest bilinear $\sigma$, i. e., it was required the spinor as an eigenspinor of the new operator, say $\mathcal{O}$ \footnote{in this context, we work in an abstract way, calling the new operators $\mathcal{A}$ and $\mathcal{B}$, defined in \cite{1305}, by $\mathcal{O}$ aiming to simplify the notation.}, entering into the dual redefinition. By assuming associativity between $\mathcal{O}$ and $S$, whose meaning is the same of Section III (except that the full Lorentz group is the relativistic group at hands), and that Eq. (\ref{eqdiracElko}) still holding, we have 
\begin{eqnarray}
(i\gamma_\mu \Lambda^{\mu}_{\beta}\partial^{' \beta}\Xi(\boldsymbol{p})\pm m)(S(\Lambda)\mathcal{O})^{-1}\mathcal{O}^{'}\lambda_\alpha^{'S/A}=0,
\end{eqnarray} which can be recast into the form 
\begin{eqnarray}
(i\gamma^{'}_{\beta}\partial^{'\beta}S(\Lambda)\Xi(\boldsymbol{p})\mathcal{O}^{-1}S^{-1}(\Lambda)\mathcal{O}^{'}\pm m S(\Lambda)(S(\Lambda)\mathcal{O})^{-1}\mathcal{O}^{'})\lambda_\alpha^{'S/A}=0.
\end{eqnarray} Therefore the following identifications immediately hold 
\begin{eqnarray}
S(\Lambda)\mathcal{O}^{-1}S^{-1}(\Lambda)\mathcal{O}^{'}=1, \label{ulti} \\ \Xi^{\prime}(\boldsymbol{p})=S(\Lambda)\Xi(\boldsymbol{p})(S(\Lambda)\mathcal{O})^{-1}\mathcal{O}^{'}. \label{ma}
\end{eqnarray} From (\ref{ulti}) one sees that the new operator must transform as  $\mathcal{O}^{'}=S(\Lambda)\mathcal{O}S^{-1}(\Lambda)$, leading to the same transformation rule of (\ref{xi1}) for the operator $\Xi(\boldsymbol{p})$. In this vein the covariance of (\ref{eqdiracElko}) is ensured. The whole covariant analysis of the corresponding bilinear quantities, then, follow immediately leading essentially to the same results.  

\section*{Acknowledgements}
JMHS thanks to CNPq (445385/2014-6; 304629/2015-4) for partial financial support. CHCV thanks to PEC-PG and RJBR thanks to CAPES for partial financial support.  

\appendix 
\section{A brief overview: the Elko formal structure}\label{apendice}
Elko spinors are eigenspinors of the charge conjugation operator $C$
\begin{eqnarray}
C\lambda^{S/A}(\boldsymbol{p}) = \pm\lambda^{S/A}(\boldsymbol{p}),
\end{eqnarray}
where
\begin{eqnarray}\label{cc}
C = \left(\begin{array}{cc}
\mathbb{O} & i\Theta \\ 
-i\Theta & \mathbb{O}
\end{array} \right)K
\end{eqnarray}
and $K$ takes the complex conjugates of any spinor that appears on the right side, while $\Theta$ is the Wigner time-reversal operator in the spin $1/2$ representation, that is given by \cite{jcap, ramond}
\begin{eqnarray}
\Theta = \left(\begin{array}{cc}
0 & -1 \\ 
1 & 0
\end{array}\right).
\end{eqnarray}
It is worth mentioning that Elko form a complete set of eigenspinors of $C$ with positive ($S$) and negative ($A$) eigenvalues. Another important aspect about these spinors is that they are constructed in such a way that carry dual helicity.

From the last considerations, the explicit form of Elko, in an arbitrary referential is
\begin{eqnarray}\label{boostElkos}
\lambda^S_{\lbrace\pm,\mp\rbrace}(\boldsymbol{p}) = \Lambda_{\pm}(p^{\mu})\lambda^S_{\lbrace\pm,\mp\rbrace}(\boldsymbol{0}),
\end{eqnarray}
and 
\begin{eqnarray}\label{boostElkoa}
\lambda^A_{\lbrace\pm,\mp\rbrace}(\boldsymbol{p}) = \Lambda_{\pm}(p^{\mu})\lambda^A_{\lbrace\pm,\mp\rbrace}(\boldsymbol{0}).
\end{eqnarray}
The action of the boost operator $\Lambda_{\pm}(p^{\mu})$ over the spinors, in the rest-frame, is given by
\begin{eqnarray}
\Lambda_{\pm}(p^{\mu})\lambda^{S/A}_{\lbrace\pm,\mp\rbrace}(\boldsymbol{0})= \underbrace{\sqrt{\frac{E+m}{2}}\bigg(1 \pm \frac{p}{E+m}\bigg)}_{boost factor}\lambda^{S/A}_{\lbrace\pm,\mp\rbrace}(\boldsymbol{0}).
\end{eqnarray}
To summarize the notation, we choose to define the boost factor by
\begin{eqnarray}
\Upsilon_{\pm} = \sqrt{\frac{E+m}{2}}\bigg(1 \pm \frac{p}{E+m}\bigg).
\end{eqnarray}
The four rest spinors are shaped by two \emph{self-conjugate}
\begin{equation}\label{Elkos}
\lambda^S_{\lbrace-,+\rbrace}(\boldsymbol{0}) = \left(\begin{array}{c}
+i\Theta[\phi^+_L(\boldsymbol{0})]^* \\ 
\phi^+_L(\boldsymbol{0})
\end{array}\right) , \qquad  \lambda^S_{\lbrace+,-\rbrace}(\boldsymbol{0}) = \left(\begin{array}{c}
+i\Theta[\phi^-_L(\boldsymbol{0})]^* \\ 
\phi^-_L(\boldsymbol{0})
\end{array}\right), 
\end{equation} 
and the other two are \emph{anti-self-conjugate}
\begin{equation}\label{Elkoa}
\lambda^A_{\lbrace-,+\rbrace}(\boldsymbol{0}) = \left(\begin{array}{c}
-i\Theta[\phi^+_L(\boldsymbol{0})]^* \\ 
\phi^+_L(\boldsymbol{0})
\end{array}\right) , \qquad  \lambda^A_{\lbrace+,-\rbrace}(\boldsymbol{0}) = \left(\begin{array}{c}
-i\Theta[\phi^-_L(\boldsymbol{0})]^* \\ 
\phi^-_L(\boldsymbol{0})
\end{array}\right).  
\end{equation}

The associated components are defined by imposing they to be eigenfunctions of helicity operator, resulting in the left-hands components given by
\begin{eqnarray}\label{compesq}
\phi_L^{+}(\boldsymbol{0}) = \sqrt{m}\left(\begin{array}{c}
\cos(\theta/2)e^{-i\phi/2} \\ 
\sin(\theta/2)e^{i\phi/2}
\end{array} \right), \qquad \phi_L^{-}(\boldsymbol{0}) = \sqrt{m}\left(\begin{array}{c}
-\sin(\theta/2)e^{-i\phi/2} \\ 
\cos(\theta/2)e^{i\phi/2}
\end{array} \right),
\end{eqnarray}
and the right-hand components presented as
\begin{eqnarray}\label{compdir}
\phi_R^{+}(\boldsymbol{0}) = \sqrt{m}\left(\begin{array}{c}
-i\sin(\theta/2)e^{-i\phi/2} \\ 
i\cos(\theta/2)e^{i\phi/2},
\end{array} \right), \qquad \phi_R^{-}(\boldsymbol{0}) = \sqrt{m}\left(\begin{array}{c}
-i\cos(\theta/2)e^{-i\phi/2} \\ 
-i\sin(\theta/2)e^{i\phi/2}
\end{array} \right).
\end{eqnarray}
Once performed the formal structure for the Elko spinors, we can now define the dual by the following general formula
\begin{eqnarray}
\stackrel{\neg}{\lambda}^{S/A}_{h}(\boldsymbol{p}) =[\Xi(\boldsymbol{p})\lambda^{S/A}_{h}(\boldsymbol{p})]^{\dag}\gamma_0,
\end{eqnarray}
with the operator $\Xi(\boldsymbol{p})$ defined as 
\begin{eqnarray}\label{XI}
\Xi(\boldsymbol{p}) \equiv \frac{1}{2m}\Big( \lambda^{S}_{\lbrace +-\rbrace}(\boldsymbol{p})\bar{\lambda}^{S}_{\lbrace +-\rbrace}(\boldsymbol{p}) + \lambda^{S}_{\lbrace -+\rbrace}(\boldsymbol{p})\bar{\lambda}^{S}_{\lbrace -+\rbrace}(\boldsymbol{p}) - \lambda^{A}_{\lbrace +-\rbrace}(\boldsymbol{p})\bar{\lambda}^{A}_{\lbrace +-\rbrace}(\boldsymbol{p})-\lambda^{A}_{\lbrace -+\rbrace}(\boldsymbol{p})\bar{\lambda}^{A}_{\lbrace -+\rbrace}(\boldsymbol{p}) \Big),
\end{eqnarray} 
in a matricial form, it reads
\begin{eqnarray}
 \Xi(\boldsymbol{p})= \left(\begin {array}{cccc} {\frac {ip\sin \left( \theta
 \right) }{m}}&{\frac {-i \left( E+p\cos \left( \theta \right) 
 \right) {{\rm e}^{-i\phi}}}{m}}&0&0\\ \noalign{\medskip}{\frac {i
 \left( E-p\cos \left( \theta \right)  \right) {{\rm e}^{i\phi}}}{m}}&
{\frac {-ip\sin \left( \theta \right) }{m}}&0&0\\ \noalign{\medskip}0&0
&{\frac {-ip\sin \left( \theta \right) }{m}}&{\frac {-i \left( E-p\cos
 \left( \theta \right)  \right) {{\rm e}^{-i\phi}}}{m}}
\\ \noalign{\medskip}0&0&{\frac {i \left( E+p\cos \left( \theta
 \right)  \right) {{\rm e}^{i\phi}}}{m}}&{\frac {ip\sin \left( \theta
 \right) }{m}}\end {array} \right), 
\end{eqnarray}
where $\Xi^{2}(\boldsymbol{p}) = \mathrm{I}$ and $\Xi^{-1}(\boldsymbol{p})$ indeed exists and is equal $\Xi(\boldsymbol{p})$ itself \cite{1305}. Now the explicit form for the dual can be readily written as 
\begin{eqnarray}
\stackrel{\neg}{\lambda}^{S/A}_{\lbrace-,+\rbrace}(\boldsymbol{p}) = +i[\lambda^{S/A}_{\lbrace+,-\rbrace}(\boldsymbol{p})]^{\dag}\gamma_0,
 \\\nonumber\\
\stackrel{\neg}{\lambda}^{S/A}_{\lbrace+,-\rbrace}(\boldsymbol{p}) = -i[\lambda^{S/A}_{\lbrace-,+\rbrace}(\boldsymbol{p})]^{\dag}\gamma_0.
\end{eqnarray} For more details about the whole construction, please see \cite{1305}. 

\section{VSR Elko spinors}\label{apendiceB}

As extensively shown in \cite{horvath} and \cite{ahluwaliahorvath}, Elko spinors can be understood as objects carrying a linear representation of SIM(2) or HOM(2) Lorentz subgroups. In this vein, it is important to explore a possible Clifford algebra basis deformation in this case, if necessary obviously. 

In order to illustrate the situation, we shall make use of the Elko spinors founded in \cite{ahluwaliahorvath}. Notice that, in general, the helicity operator does not commute with the VSR boost. Therefore, one cannot freely chose the rest spinors as a basis for such an operator. This is the kernel of the subtle difference between the Elko spinors studied along the main text and the spinors here investigated. 

For an arbitrary momentum, it is possible to see that in the VSR scope we have \cite{ahluwaliahorvath}
\begin{eqnarray}\label{vixi}
\chi^S_{\{-,+\}}(\boldsymbol{p}) = \sqrt{m}\left( \begin{array}{c}
i\frac{p_x-ip_y}{\sqrt{m(p_0-p_z)}}e^{i\phi/2} \\ 
i\sqrt{\frac{p_0-p_z}{m}}e^{i\phi/2} \\ 
\sqrt{\frac{p_0-p_z}{m}}e^{-i\phi/2} \\ 
-\frac{p_x+ip_y}{\sqrt{m(p_0-p_z)}}e^{-i\phi/2}
\end{array} \right),
\end{eqnarray} as a prototype spinor and 
\begin{eqnarray}
\stackrel{\neg}{\chi}^S_{\{-,+\}}(\boldsymbol{p}) = \sqrt{m}\left(\begin{array}{cccc}
0 & -i\frac{e^{-i\phi/2}}{\sqrt{\frac{p_0-p_z}{m}}} & \frac{e^{i\phi/2}}{\sqrt{\frac{p_0-p_z}{m}}} & 0
\end{array} \right),
\end{eqnarray} as its dual, whose construction obeys the same previous prescription, i. e., $\chi^{S/A}_{h}(\boldsymbol{p})=[\Xi(\boldsymbol{p})_{VSR}\chi^{S/A}_{h}(\boldsymbol{p})]^{\dag}\gamma_0$. All the aspects investigated in the main text have a parallel here. The only differences are the general boost of VSR, which reads 
\begin{eqnarray}\label{lalala}
\mathcal{V}_{VSR} = \left(\begin{array}{cccc}
\sqrt{\frac{m}{p_0-p_z}} & \frac{p_1-ip_2}{\sqrt{m(p_0-p_3)}} & 0 & 0 \\ 
0 & \sqrt{\frac{p_0-p_z}{m}} & 0 & 0 \\ 
0 & 0 & \sqrt{\frac{p_0-p_z}{m}} & 0 \\ 
0 & 0 & -\frac{p_1+ip_2}{\sqrt{m(p_0-p_3)}} & \sqrt{\frac{m}{p_0-p_z}}
\end{array}  \right),
\end{eqnarray} and the $\Xi(\boldsymbol{p})_{VSR}$ operator. In order to find this last operator, it is possible to see that starting from Eq. (\ref{XI}), but replacing the usual spinors by VSR Elko spinors (just as (\ref{vixi})) in the composition of $\Xi(\boldsymbol{p})_{VSR}$, one arrive at  
\begin{eqnarray}
\Xi(\boldsymbol{p})_{VSR}= \left( \begin {array}{cccc} {\frac {i \left( p_{{x}}-ip_{{y}}
 \right)e^{i\phi}}{m}}&{\frac {-i \left( p_{{x}}-ip_{{y
}} \right) ^{2}{{\rm e}^{i\phi}}}{m \left( p_{{0}}-p_{{z}} \right) }}-
{\frac {im{{\rm e}^{-i\phi}}}{p_{{0}}-p_{{z}}}}&0&0
\\ \noalign{\medskip}{\frac {i \left( p_{{0}}-p_{{z}} \right) {{\rm e}
^{i\phi}}}{m}}&{-\frac {i \left( p_{{x}}-ip_{{y}}
 \right)e^{i\phi}}{m}}&0&0\\ \noalign{\medskip}0&0&{-\frac {i \left( p_{{x}}+ip_{{y}}
 \right)e^{-i\phi}}{m}}&{\frac {-i \left( p_{
{0}}-p_{{z}} \right) {{\rm e}^{-i\phi}}}{m}}\\ \noalign{\medskip}0&0&{
\frac {i \left( p_{{x}}+ip_{{y}} \right) ^{2}{{\rm e}^{-i\phi}}}{m
 \left( p_{{0}}-p_{{z}} \right) }}+{\frac {i{{\rm e}^{i\phi}}m}{p_{{0}
}-p_{{z}}}}&{\frac {i \left( p_{{x}}+ip_{{y}}
 \right)e^{-i\phi}}{m}}\end {array} \right).
\end{eqnarray} This operator acts in VSR spinors just as the $\Xi(\boldsymbol{p})$ operator does in usual Elko spinors: it change the spinor type, obeys $\Xi^{2}(\boldsymbol{p})_{VSR}=1$ and have determinant equal to one, ensuring the existence of the inverse. 

With such ingredients, it is possible to see that all the previous steps are repeated here, namely: the FPK identities are not respected, except if the Clifford basis undergoes the same deformation (with $\Xi(\boldsymbol{p})_{VSR}$ replacing $\Xi(\boldsymbol{p})$); $\sigma_{VSR}$ and $K^{\mu}_{VSR}$ are non null real quantities, $\omega_{VSR}$ is zero, and the remain bilinears are imaginary. Besides, as the relation (\ref{eqdiracElko}) still holding, the covariance structure is also the same, provided (\ref{lalala}) instead of the usual Lorentz transformation.

\end{document}